# AN OVERVIEW OF MOBILE AD HOC NETWORKS FOR THE EXISTING PROTOCOLS AND APPLICATIONS

Saleh Ali K.Al-Omari[1], Putra Sumari[2]

School of Computer Science, Universiti Sains Malaysia, 11800 Penang, Malaysia
Salehalomari2005@yahoo.com , putras@cs.usm.my

## ABSTRACT

*Mobile Ad Hoc Network (MANET) is a collection of two or more devices or nodes or terminals with wireless communications and networking capability that communicate with each other without the aid of any centralized administrator also the wireless nodes that can dynamically form a network to exchange information without using any existing fixed network infrastructure. And it's an autonomous system in which mobile hosts connected by wireless links are free to be dynamically and some time act as routers at the same time, and we discuss in this paper the distinct characteristics of traditional wired networks, including network configuration may change at any time , there is no direction or limit the movement and so on, and thus needed a new optional path Agreement (Routing Protocol) to identify nodes for these actions communicate with each other path, An ideal choice way the agreement should not only be able to find the right path, and the Ad Hoc Network must be able to adapt to changing network of this type at any time. and we talk in details in this paper all the information of Mobile Ad Hoc Network which include the History of ad hoc, wireless ad hoc, wireless mobile approaches and types of mobile ad Hoc networks, and then we present more than 13 types of the routing Ad Hoc Networks protocols have been proposed. In this paper, the more representative of routing protocols, analysis of individual characteristics and advantages and disadvantages to collate and compare, and present the all applications or the Possible Service of Ad Hoc Networks*

## 1. INTRODUCTION

With the widespread rapid development of computers and the wireless communication, the mobile computing has already become the field of computer communications in high-profile link. Mobile Ad Hoc Network (MANET) is a completely wireless connectivity through the nodes constructed by the actions of the network, which usually has a dynamic shape and a limited bandwidth and other features, network members may be inside the laptop, Personal Digital Assistant (PDA), mobile phones, MP3 players, and digital cameras and so on. On the Internet, the original Mobility (mobility) is the term used to denote actions hosts roaming in a different domain; they can retain their own fixed IP address, without the need to constantly changing, which is Mobile IP technology. Mobile IP nodes in the main action is to deal with IP address management, by Home Agent and Foreign Agent to the Mobile Node to packet Tunnelling, the Routing and fixed networks are no different from the original; however, Ad Hoc Network to be provided by Mobility is a fully wireless, can be any mobile network infrastructure, without a base station, all the nodes can be any link, each node at the same time take Router work with the Mobile IP completely different levels of Mobility. Early use of the military on the Mobile Packet Radio Networked in fact can be considered the predecessor of MANET, with the IC technology advances, when the high-tech communication equipment, the size, weight continuously decreases, power consumption is getting low, Personal Communication System (Personal Communication System, PCS) concept evolved, from the past few years the rapid popularization of mobile phones can be seen to communicate with others anytime, anywhere, get the latest information, or exchange the required information is no longer a dream, And we have gradually become an integral part of life. Military purposes, as is often considerable danger in field environment, some of the major basic communication facilities, such as base stations, may not be available, in this case, different units, or if you want to communicate between the forces, we must rely on This cannot MANET network infrastructure limitations. In emergency relief, the mountain search and rescue operations at sea, or even have any infrastructure can not be expected to comply with the topographical constraints and the pressure of time under the pressure, Ad Hoc Network completely wireless and can be any mobile feature is especially suited to disaster relief operations. When personal communication devices and more powerful, some assembly occasions, if you need to exchange large amounts of data, whether the transmission of computer files or applications that display. if we can link into a temporary network structure, then the data transmission will be more efficient without the need for large-scale projection equipment would not have point to point link equipment (such as network line or transmission line). The current wireless LAN technology, Bluetooth is has attracted considerable attention as a development plan. Bluetooth's goal is to enable wireless devices to contact with each other, if the adding the design of Ad Hoc Network (MANET).

## 2. RELATED BACKGROUND

Nowadays, the information technology will be mainly based on wireless technology, the conventional mobile network and cellular are still, in some sense, limited by their need for infrastructure for instance base station, routers and so on. For the Mobile Ad Hoc Network, this final limitation is eliminated, and the Ad Hoc Network are the key in the evolution of wireless network and the Ad Hoc Network are typically composed of equal node which communication over wireless link without any central control. Although military tactical communication is still considered as the primary application for Ad Hoc Networks and commercial interest in this type of networks continues to grow. And all the applications such as rescue mission in time of natural disasters, law enforcement operation, and commercial as rescue and in the sensor network are few commercial examples, but in this time it's become very important in our life and they become use it.

The Ad Hoc Networking application is not new one and the original can be traced back to the Defence Advanced Research Projects Agency (DARPA) Packet Radio Networking (PRNET) project in 1972[1, 2, 3] which evolved into the survivable adaptive radio networks (SURAN) program [4] .which was primarily inspired by the efficiency of the packet switching technology for instance the store/forward routing and then bandwidth sharing, it's possible application in the mobile Ah Hoc Networks environments. as well as in the Packet Radio Networking devises like Repeaters and Routers and so on, were all mobile although mobility was so limited in that time, theses advanced protocol was consider good in the 1970s.after few years advance in Micro Electronics technology and it's was possible to integrate all the nodes and also the network devices into a single unit Called Ad Hoc Nodes. And then the advance such as the flexibility, resilience also mobility and independence of fixed infrastructure, and in that time they so interesting to use it immediately among military battlefield, Ad hoc networks have played an important role in military applications and related research efforts, for example, the global mobile information systems (GloMo) program [5] and the near-term digital radio (NTDR) [6] program. And also has been the increase in the police, commercial sector and rescue agencies in use of such networks under disorganized environments.

Ad Hoc network research stayed long time in the realm of the military. And in the middle of 1990s with advice of commercial radio technology and the wireless became aware of the great advantages of Mobile Ad Hoc networks outside the military battlefield domain, and then became so active research work on ad hoc network start in 1995 in the conference session of the Internet Engineering Task Force (IETF) [7]. And then in 1996 this works had evolved into Mobile Ad Hoc Network, in that time focused to discussion cantered in military satellite network, wearable computer network and tactical network with specific concerns begin raised relative to adaptation of existing routing protocols to support IP network in dynamic environments, as well as they make the charter of the Mobile Ad Hoc Network Working Group (MANETWG) of the Internet Engineering Task Force (IETF) also the work inside the MANTs relies on other existing IETF standard such as Mobile IP and IP addressing  and most of the currently available solutions are not designed to scale to more than a few hundred nodes. Currently, the research in Mobile Ad Hoc Network became so active and vibrant area and the efforts this research community together with the current and future (MANET) enabling radio technology.

Recently, the Ad Hoc Wireless Network and computing consortium was established with the aim to coalescing the interests and efforts to use it anywhere such as academic area and industry and so on. And in order to apply this technology to application ranging for the Home wireless to wide area peer to remote networking and communications. And it's will certainly pave the way for commercially viable MANET networks and their new and exciting applications, which began to appear in all fields in this life.

More recently, the computer has became spread significantly in the all the place and after a pervasive computing environment can be expected based on the recent progresses and advances in computing and communication technologies. Next generation of mobile communications will include both prestigious infrastructure wireless networks and novel infrastructureless mobile ad hoc networks (MANETs).

## 3. WIRELESS AD HOC NETWORKS

MANET is a collection of two or more devices or nodes or terminals with wireless communications and networking capability that communicate with each other without the aid of any centralized administrator also the wireless nodes that can dynamically form a network to exchange information without using any existing fixed network infrastructure. And it's an autonomous system in which mobile hosts connected by wireless links are free to be dynamically and some time act as routers at the same time.  All nodes in a wireless ad hoc network act as a router and host as well as the network topology is in dynamically, because the connectivity between the nodes may vary with time due to some of the node departures and new node arrivals. The special features of Mobile Ad Hoc Network (MANET) bring this technology great opportunity together with severe challenges [8].All the nodes or devises responsible to organize themselves dynamically the communication between the each other and to provide the necessary network functionality in the absence of fixed infrastructure or we can call it ventral administration, It implies that maintenance, routing and management, etc. have to be done between all the nodes. This case Called Peer level Multi Hopping and that is the main building block for Ad Hoc Network. In the end, conclude that the Ad Hoc Nodes or devices are difficult and more complex than other wireless networks. Therefore, Ad Hoc Networks form sort of clusters to the effective implementation of such a complex process. In the following figure 1 will shows some nodes forming ad hoc networks, and there are some nodes more randomly in different direction and different speeds.

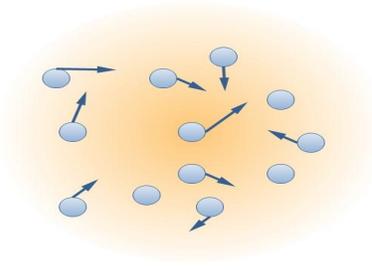

Figure 1. Ad Hoc Network: Nodes mover randomly in different direction and different speeds

In the past few years, the people became realized to use all the technology so widely and the people's future living environments are emerging, based on information resource provided by the connections of different communication networks for clients also we have seen a rapid expansion in the field of Mobile Computing because  the proliferation not expensive, widely available wireless devices .A new small devices such as personal communication like cell phones, laptops, Personal Digital Assistants (PDAs),handhelds, and also there's a lot of traditional home appliances such as a digital cameras, cooking ovens, washing machines, refrigerators and thermostats, with computing and communicating powers attached. Expand this area to became a fully pervasive and so widely. With all of this, the technologies must be formed the good and new standard of pervasive computing, that including the new standards, new tools, services, devices, protocols and a new architectures.

As well as the people in this time, or the users of internet users in Ad Hoc network through increase in the use of its advantage is that not involve any connection link and the wiring needed to save space, and building low cost, and improve the use, and can be used in mobile phone, because of these advantage local wireless network architecture readily. And also beads in these advantages the wireless network can be used in the local area network terminal part of the wireless [9].

## 4. Wireless Mobile Approaches

The past decade the Mobile Network is the only one much important computational techniques to support computing and widespread, also advances in both software techniques and the hardware techniques have resulted in mobile hosts and wireless networking common and miscellaneous. Now we will discuss about to distinct approaches very important to enabling Mobile wireless Network or IEEE 802.11 to make a communication between each other [10, 11]. Firstly infrastructure wireless networks and secondly, infrastructureless wireless networks (Ad Hoc Networks) and we will clarify both in bottom.

### 4.1. Infrastructure Wireless Networks

In this architecture that allow the wireless station to make a communication between each other, and this type relies on the third fixed party and we call it a Base Station, as shows in this figure 2, and that will handover the offered traffic from the Station to another, the same entity will regulate or organize the allocation of radio resources. When a source node likes to communicate with a destination node, the former notifies the base station. At this point, the communicating nodes do not need to know anything about the route from one to another. All that matters is that the both the source and the destination nodes are within the transmission range for the Base Station and then if there's any one  loses this condition, the communication will frustration or abort.

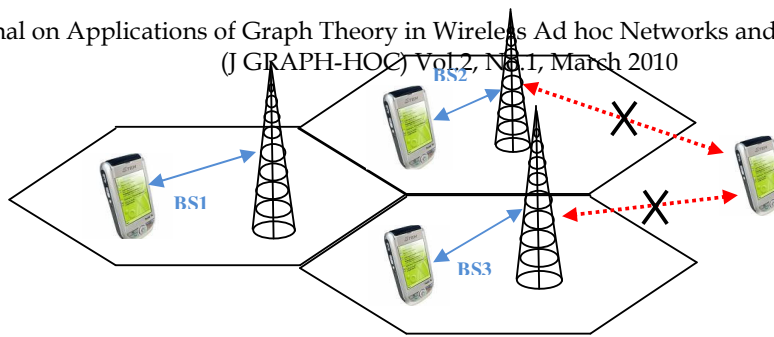

Figure .2. Shows of the infrastructure network

## 4.2 Infrastructureless Wireless Networks:

The mobile wireless network As is well known a Ad Hoc Network MANETs, As has been previously defined in the Bidder is a collection of two or more devices or nodes or terminals with wireless communications and networking capability that communicate with each other without the aid of any centralized administrator also the wireless nodes that can dynamically form a network to exchange information without using any existing fixed network infrastructure. And it's an autonomous system in which mobile hosts connected by wireless links are free to be dynamically and some time act as routers at the same time [12, 13]. the infrastructureless it's important approaches in this technique to communication technology that supports truly pervasive computing widely duo to there's a lot of context information need to exchange between mobile units can not rely on the fixed network infrastructure, but in this time the communication wireless became develops very fast.

In figure 3 we will see a small example for the Ad Hoc networks, to explain the work for the Ad Hoc network.

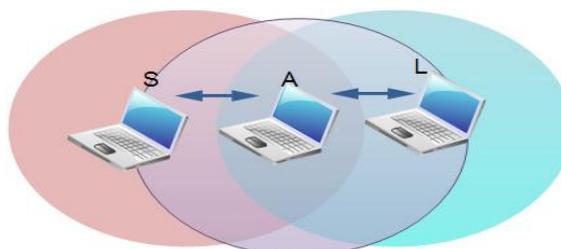

Figure 3. Illustration of the infrastructureless networks (Ad Hoc Networks)

This figure illustrates the modus operandi of Ad Hoc networks, there's a three nodes Ad Hoc Network (S, A, L), the source node (S) need to make a communication with the destination node (L) and both of them (S, L) not in the same transmission range of each others, here both they must use the node (A)to send/ receive or forewords the Packets from source to the destination that means from node to another node.(R) is a node work as host and router in the same time .
Also as we know the definition for the router is an entity that determines the path to be used in order to forward a packet towards its last destination. And then the router chooses the next node to which a packet should be forwarded according to its current understanding of the state of the network.

## 5. Types of Ad hoc network

The wireless Ad Hoc Network divided into two main types, firstly quasi-static Ad Hoc Network secondly, Mobile Ad Hoc Network (MANET). In the quasi-static Ad Hoc network the nodes may be portable or static, because the power controls and link failures, the resulting network topology may be so active. The Sensor Network is an example for the quasi-static Ad Hoc Network [14]. In the Mobile Ad Hoc network (MANET) here the entire network may be mobile and the nodes may move fast relative to each other. And now we will discuss both of them.

### 5.1 Mobile Ad Hoc Networking (MANET)

Mobile Ad hoc Networking (MANET) is a group of independent network mobile devices that are connected over various wireless links. It is relatively working on a constrained bandwidth. The network topologies are dynamic and may vary from time to time. Each device must act as a router for transferring any traffic among each other. This network can operate by itself or incorporate into large area network (LAN).

There are three types of MANET. It includes Vehicular Ad hoc Networks (VANETs), Intelligent Vehicular Ad hoc Networks (InVANETs) and Internet Based Mobile Ad hock Networks (iMANET). The set of application for MANETs can be ranged from small, static networks that are limited by power sources, to large-scale, mobile, highly dynamic networks. On top of that, the design of network protocols for these types of networks is face with multifaceted issue. Apart from of the application, MANETs need well-organized distributed algorithms to determine network organization, link scheduling, and routing. Conventional routing will not work in this distributed environment because this network topology can change at any point of time. Therefore, we need some sophisticated routing algorithms that take into consideration this important issue (mobile network topology) into account. While the shortest path (based on a given cost function) from a source to a destination in a static network is usually the optimal route, this idea is not easily far-reaching to MANETs. Some of the factors that have become the core issues in routing include variable wireless link quality, propagation path loss, fading, interference; power consumed, and network topological changes. This kind of condition is being provoked in a military environment, beside these issues in routing, we also need to guarantee assets security, latency, reliability, protection against intentional jamming, and recovery from failure. Failing to abide to of any of these requirements may downgrade the performance and the dependability of the network.

### 5.2 Mobile Ad Hoc Sensor Network

A mobile ad-hoc sensor network follows a broader sequence of operational, and needs a less complex setup procedure compared to typical sensor networks, which communicate directly with the centralized controller. A mobile ad-hoc sensor or Hybrid Ad Hoc Network includes a number of sensor spreads in a large geographical area. Each sensor is proficient in handling mobile communication and has some level of intelligence to process signals and to transmit data. In order to support routed communications between two mobile nodes, the routing protocol determines the node connectivity and routes packets accordingly. This condition has makes a mobile ad-hoc sensor network highly flexible so that it can be deployed in almost all environments [15].

The Wireless ad-hoc sensor networks [16] are now getting in style to researchers. This is due to the new features of these networks were either unknown or at least not systematized in the past. There are many benefits of this network, it includes:

- Use to build a large-scale networks
- Implementing sophisticated protocols
- Reduce the amount of communication (wireless) required to perform tasks by distributed and/or local precipitations.
- Implementation of complex power saving modes of operation depending on the environment and the state of the network.

With the above-mentioned advances in sensor network technology, functional applications of wireless sensor networks increasingly continue to surface. Examples include the replacement of existing detecting scheme for forest fires around the world. Using sensor networks, the detecting time can be reduced significantly. Secondly is the application in the large buildings that at present use various environmental sensors and complex control system to execute the wired sensor networks. In a mobile ad-hoc sensor networks, each host may be equipped with a variety of sensors that can be organized to detect different local events. Besides, an ad-hoc sensor network requires a low setup and administration costs [16, 17, 18].

## 6. The Traffic Types in the Ad Hoc Networks

The Traffic Types in the Ad Hoc Networks are so different from the infrastructure wireless network, and then now we will see these types. The first one Peer to Peer (P2P) the second remote to remote and lastly dynamic traffic. So now we will discuss every one [19].

**Firstly,** Peer to peer: communication between two nodes in the same area, that means which are within one hop. Network traffic (in bits per second) is usually fixed. **Secondly**, remote to remote: Communication between two nodes beyond a single hop, but maintain a stable route between them. This may be the result of a number of Nodes, to stay within the range of each other in one area or may move as a group. Movement it's a similar to the standard network traffic. Finally, Dynamic traffic: its will happened when the nodes are movie dynamically around and then the routers must be reconstructed. This results in a poor connectivity and network activity in short bursts. For example in IEEE 802.11 network and the basic structure divided into two types firstly infrastructures wireless LAN, the second structure Ad Hoc Wireless LAN.

### 6.1 Infrastructures Wireless LAN

In this kind of network as we shows in the figure 4, the network in any architecture will be an access point; its function is one or more of the wireless local area network and the existing cable network systems to link, so that stations within the wireless local area network and external nodes can connect with each other. It is characterized by a fixed and pre-positioning a good base station location, the static backbone network topology, a good environment and a stable connection, the base station that is doing a good job when you set up detailed plans [60].

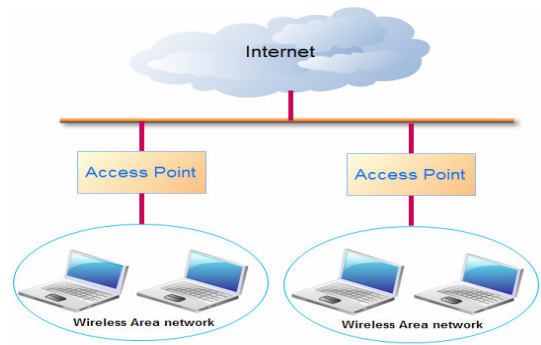

Figure 4. Infrastructure wireless LAN Architecture.

### 6.2 Ad Hoc Wireless LAN

That means it's without or relies on infrastructures wireless local area network, which only targeted at local area network within the framework of each machine is able to be linked up into networks, regardless of whether the communication with the outside world, then such a structure, either one or two users can communicate directly with each other, and this structure is composed of at least composed of two or more workstations. Is characterized by no fixed base stations, network will be rapidly changing; dynamic network topology is vulnerable to interference, to automatically form a network without infrastructure and adapt to topology changes. For more explain shows the figure 5 for Ad Hoc Wireless network.

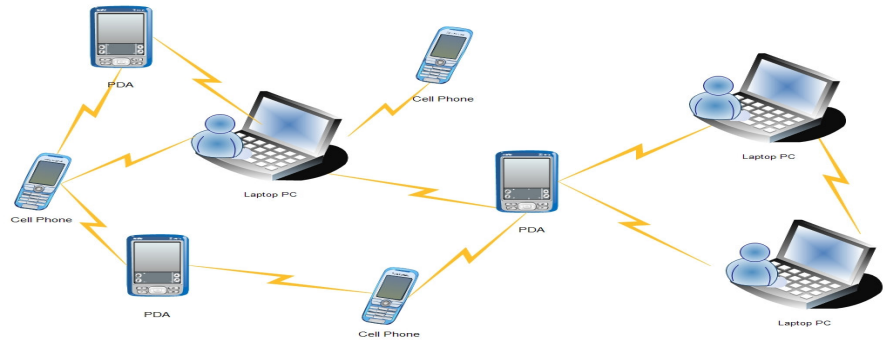

Figure 5. Wireless Ad Hoc Network

## 7. The fundamentals for the Mobile Ad Hoc Networks:

According for example to IETF RFC 2501, MANET has characteristics can be divided into the following sections:

**7.1** Dynamic topologies: nodes can move freely, network topology may change rapidly, restructuring, but also may also have symmetric and asymmetric links.

**7.2** Bandwidth-constrained, variable capacity links Compared with the wired network environment, the capacity of the wireless link itself is relatively small, but also susceptible to external noise, interference, and signal attenuation effects.

**7.3** Energy-constrained operation A laptop or handheld computers are often used batteries to provide power, how to save electricity in the context of depletion of system design is also necessary to consider the point.

**7.4** Limited physical security Network Security With the network deeply embedded in our daily lives and the benefits have become increasingly important in the wireless network to provide security support is also an important issue.

## 8. Ad Hoc Network Routing Protocol Performance Issues

As the MANET with the traditional wired, fixed networks have many different characteristics, so to design a suitable routing protocol for MANET operating environment must also consider the different directions, the following sub-qualitative and quantitative aspects of the discussion:

**8.1 on the qualitative aspects, can be divided into**

**8.1.1** *Distribution operation*: due to the existence of MANET where there is no prerequisite for the construction of the underlying network, so routing can not rely on a particular node to operate.

**8.1.2** *Loop-freedom*: all the routing protocol should be consistent with the characteristics; we must ensure the normal work in order to avoid waste of bandwidth.

**8.1.3** *Demand-based operation*: In order to reduce the burden on each node, if the link is not so much the demand should be considered when using On-demand approach to the establishment of the path, and only when the need for a particular path query, the establishment of the path.

**8.1.4** *Proactive operation*: with the On-demand concept of the contrary, if the network resources fairly adequate, proactive table-driven approach could speed up the path to the establishment of speed.

**8.1.5** *Security*: Because it is the wireless environment, to how to ensure the security of the connection can not be ignored will be part of network security is also a MANET from theory to implementation of the key challenges

**8.1.6** *Sleep Period operation*: As the MANET nodes are generally smaller wireless devices, using the battery as a power supply, how to save power consumption, or for no work, the node goes into sleep mode, can operate more smoothly so that MANET. Also the nodes of a Mobile Ad Hoc Networks may stop transmitting or receiving or both, also even receiving requires power for arbitrary time periods and the routing protocol should be able to accommodate such sleep periods without overly adverse consequences. This property may require close coupling with the link-layer protocol through a standardized interface.

**8.1.7** *Unidirectional link support:* MANET nodes have heterogeneous characteristics, and some may be notebook computers, and some may be PDA, and some might be even smaller devices, in this environment, the transmission characteristics of asymmetric more significant than the wired environment.

## 8.2 On the quantity, can be divided into

**8.2.1** *End-to-end data throughput and delay*: data transmission rate and delay in the case that every routing protocol must take into account the focus should be how to find the best path? Is the maximum bandwidth or minimum latency, or the link to the most stable? Considered more likely to make more complicated routing protocol, but it is possible to significantly improve the transmission quality.

**8.2.2** *Route Acquisition time:* While the table-driven generally higher than on-demand performance good, but many of the former to pay the price, which, if properly designed, for example, there is more commonly used in the path cache, or a certain fixed path , can improve the path to the establishment of time.

**8.2.3** *Percentage Out-of-order delivery:* real-time data for this part of the more stringent requirements, and general information will not affect how and upper TCP cooperation is also IP routing work.

**8.2.4** *Efficiency:* the simplest method, the smallest control overhead done the most complete, most powerful feature is a common goal for all routing protocol.

## 9. Types of Ad Hoc Protocols

Ad Hoc Network routing protocols is divided to three type of routing protocols, which that depending on a different of routing protocols [20-26].

### 9.1 Oriented routing table (Table-driven)

It is an active routing environment in which the intervals between the wireless nodes will send medical information with more paths.  Each wireless node is on the basis of information gathered recently to change its route table. When the network topology change makes the original path is invalid, or the establishment of any new path, all nodes will receive updates on the status path. The path will be continuously updated, so that the node in time of peace on its own routing tables is ready, and immediately available when needed. However, such agreements must be periodically to broadcast messages, so a considerable waste of wireless bandwidth and wireless node power, but if you want to reduce the broadcast bandwidth consumption caused by a large number, we should lengthen the interval between each broadcast time, which in turn will result in the path table does not accurately reflect network topology changes.

### 9.2 Demand-driven (On-demand)

When needed to send packets only it began to prepare to send the routing table. When a wireless node needs to send data to another wireless node, the source client node will call a path discovery process, and stored in the registers of this path. The path is not valid until the expiration or the occurrence of conditions of the agreement with the first phase of a ratio of such agreements in each node. A smaller amount of data needed, and do not need to save the entire network environment and the routing information. The main benefit of this agreement is that the use of a lower bandwidth, but the drawback is that not every wireless node that sends packets can always quickly find the path. The path discovery procedure can cause delays and the average delay time is longer [27].

### 9.3 Hybrid

It is an improvement of the abovementioned two, or the combination of other equipment, such as global positioning system (GPS) and other equipment, participate in the study of mechanisms to facilitate the routing of the quick search, and data transmission.[28,29]

However, there are already more than 13 kinds of the above routing protocol have been proposed, following the more representative for several separate presentations, and to compare their individual differences lie. And then we will discuss about everyone and we will show the way to works everyone works.

## 10.    The compare between Proactive versus Reactive and then Clustering and Hierarchical Routing

### 10.1 Proactive versus Reactive Approaches

Ad hoc routing protocols can be classified as into two types; proactive or On Demand (reactive) base on each own strategy [56]. Proactive protocols demand nodes in a wireless ad hoc network to keep track of routes to all possible destinations. This is important because, whenever a packet requests to be forwarded, the route is beforehand identified and can be used straight away. Whenever there's modification in the topology, it will be disseminated throughout the entire network. Instances include "destination-sequenced distance-vector" (DSDV) routing [30], "wireless routing protocol" (WRP) [32], "global state routing" (GSR) [57], and "fisheye state routing" (FSR) [58] and in next section will discuss about everyone.

On-demand (reactive) protocols will build the routes when required by the source node, in order for the network topology to be detected as needed (on-demand). When a node needs to send packets to several destinations but has no routes to the destination, it will start a route detection process within the network. When a route is recognized, it will be sustained by a route maintenance procedure until the destination becomes unreachable or till the route is not wanted anymore. Instances include "ad hoc on-demand distance vector routing" (AODV) [33], "dynamic source routing" (DSR) [37], and "Cluster Based Routing protocol" (CBRP) [59]. Proactive protocols comprise the benefit that new communications with arbitrary destinations experience minimal delay, but experience the disadvantage of the extra control overhead to update routing information at all nodes. To overcome with this limitation, reactive protocols take on the opposite method by tracking down route to a destination only when required. Reactive protocols regularly utilize less bandwidth compared to proactive protocols, however it is a time consuming process for any route tracking activity to a destination proceeding to the authentic communication. Whenever reactive routing protocols must relay route requests, it will create unnecessary traffic if route discovery is required regularly.

### 10.2 Clustering and Hierarchical Routing

Scalability is one of the major tribulations in ad hoc networking. The term scalability in ad hoc networks can be defined as the network's capability to provide an acceptable level of service to packets even in the presence of a great number of nodes in the network. If the number of nodes in the network multiply for proactive routing protocols, the number of topology control messages will increases nonlinearly and it will use up a large fraction of the available bandwidth.

While in reactive routing protocols, if there are a large numbers of route requests propagated to the entire network, it may eventually become packet broadcast storms. Normally, whenever the network size expands beyond certain thresholds, the computation and storage requirements become infeasible. At a time whenever mobility is being taken into consideration, the regularity of routing information updates may be extensively enhanced, and will deteriorate the scalability issues.

In order to overcome these obstacles and to generate scalable and resourceful solutions, the solution is to use hierarchical routing. Wireless hierarchical routing is based on the idea of systematizing nodes in groups and then assigns the nodes with different task within and outside a group. Both the routing table size and update packet size are decreased by comprising only a fraction of the network. For reactive protocols, restricting the scope of route request broadcasts can assists in improving the competency. The best method of building hierarchy is to gather all nodes geographically near to each other into groups. Every cluster has a principal node (cluster head) that corresponds with other nodes. Instances of hierarchical ad hoc routing protocols include "zone routing protocol" (ZRP) [46].

## 11.     Existing Ad Hoc Protocols

For the Ad Hoc network there are more than 13 kinds of the above routing protocol have been proposed, following the more representative for several separate presentation, and to compare between them, and for more dilates about existing ad hoc network protocols [62].

### 11.1 Destination-Sequenced Distance-Vector Routing (DSDV)

Destination-Sequenced Distance-Vector Routing [30, 61] is based on traditional Bellman-Ford routing algorithms were developed by the improvement, and a routing table-based protocol. Each node in an operation must be stored a routing table, which records all the possible links with the nodes in the node and the distance like the number of hops, routing table within each record also contains a sequence number, which is used to determine are there any more old path in order to avoid routing table generation. DSDV is basically on the Internet Distance-Vector Routing the same, but more destination sequence number of the record, makes the Distance-Vector Routing more in line with this dynamic network MANET needs, In addition, when network topology changes are less frequent when the routing table does not need to exchange all the information, DSDV, within each node, together with a table, is used to record the routing table changes from the last part of the exchange so far, if you change a lot of the conduct of all the information The exchange, known as the full dump packets, if the change very little, it is only for the part of the exchange, known as the incremental packet.

### 11.2 Global State Routing (GSR)

Global State Routing (GSR) [58] is almost the same as DSDV, because it has the idea of link state routing but it makes a progress by decreasing the flooding of routing messages. In this algorithm, each node maintains a neighbour list, a topology table, a next hop table and a distance table.
•The neighbour list of a node includes the list of its neighbours (all nodes that can be heard by it).
• The link state information for each destination is maintained in the topology table together with the timestamp of the information.
• The next hop table includes the next hop to which the packets for each destination must be dispatched.
• The distance table contains the shortest distance to each destination node. The routing messages will be created on a link change as in all link state protocols. Whenever it accepts a routing message, the node updates its topology table if the sequence number of the message is later than the sequence number stored in the table and it then reconstructs its routing table and broadcasts the information to its neighbours.

### 11.3 Cluster head Gateway Switch Routing (CGSR)

Cluster head Gateway Switch Routing [31] is to build from the DSDV above a routing protocol, using a cluster head to manage a group of action nodes, that is, the action is divided into a group of a group of nodes, each elected by a head, the cluster head among through a gateway to connect to each other, into a hierarchical structure. Whether a link between nodes within a cluster, or a link between each cluster head, are based on DSDV routing, so each node also needs a routing table for the record, in addition to DSDV need some information is also necessary to routing table with a record of all the other nodes and the corresponding cluster head.

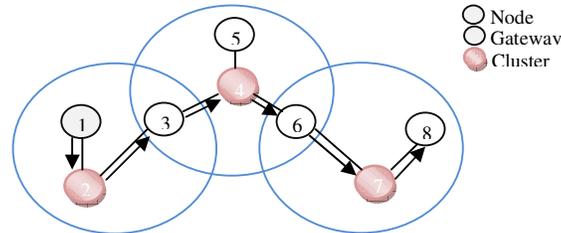

Figure 6. CGSR: routing from node 1 to node 8

In the end web of which nodes should be selected as cluster head, and when the cluster head movement and also how to avoid future path changes significantly under to find a new head, all in real CGSR to allow for the more difficult areas.

### 11.4 Wireless Routing Protocol (WRP)

Wireless Routing Protocol [32] makes use of the routing table at each node in the record to complete the routing, and DSDV with CGSR difference is that, WRP require each node to operate a record four tables, namely Distance table, Routing table, Link-cost table, Message retransmission list table. WRP use the update message between adjacent nodes in each pass is used to determine whether the adjacent nodes to maintain their link relationship, and Message retransmission list (MRL) is used to update records which need to re-transmission, and which update needs acknowledgement. WRP use of distance and the second-to-last hop information to find the path, such an approach can effectively improve the distance-vector routing possible count-to-infinity problem.

### 11.5 Fisheye State Routing (FSR)

Fisheye State Routing (FSR) [59] is an enhancement of GSR. The large size of update messages in GSR dissipates a substantial amount of network bandwidth. In order to overcome this problem, FSR will use a method where each updated messages would not includes information about all nodes. As an alternative, it swaps information about neighbouring nodes regularly than it does about farther nodes, thus reducing the update message size. In this way, each node gets accurate information about near neighbours and accuracy of information decreases as the distance from the node increases. Even though a node does not have accurate information about distant nodes, the packets are routed correctly because the route information becomes more and more accurate as the packet moves closer to the destination.

### 11.6 Ad Hoc On-Demand Distance Vector Routing (AODV)

Ad Hoc On-Demand Distance Vector Routing using distance-vector concept [33,34,35], but in several different ways and the above is that, AODV does not maintain a routing table, but when a node needs to communicate with another node on demand only to the approach to building routing table. When a node wants to send data to another node in the network, the first to broadcast a Route Request (RREQ) packet [36], RREQ where the record that this is given by which a source is to be used to find which of a destination node. RREQ in the network is a kind of flooding of the transfer mode, destination until they were received, of course, a node can only be processed once on the same RREQ in order to avoid routing loop generation. In theory all the nodes between the source and the destination of the RREQ will be passing a temporary record will be on the last hop of the RREQ via Path of information, when the destination of the RREQ received from different places, choose a shortest path, and to the source sent the direction of
Route Reply (RREP).

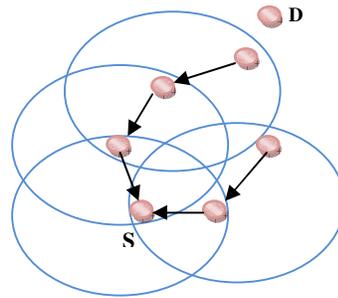

Figure 7 AODV: reverse path formation

As the RREP of passing along the nodes on this path will be a record of the relevant
information, when the RREP was sent to a sent RREQ the source the beginning, this section of
the path from source to destination even been established, and thereafter source can use this
route to send packets to the destination.

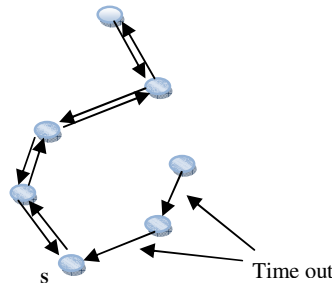

Figure 8. AODV: forward path formation

It is noteworthy that, AODV and DSDV routing table inside the same have also recorded a
destination sequence number, to avoid routing loop from occurring, but also to ensure the path
recorded in the latest expression of the topology.

**11.7 Dynamic Source Routing (DSR)**

Dynamic Source Routing [37,38,39] As the name suggests is the use of the concept of source
routing, the routing information that is directly recorded in the inside of each packet, but to be in
the MANET environment, the use of such a special, DSR is needed only when the path to find
out the path, that is, on -demand.

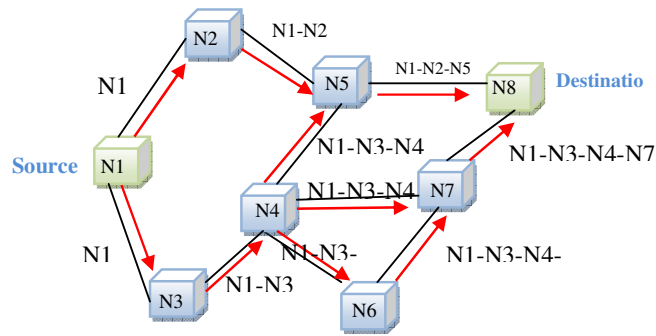

Figure 9. DSR: route request

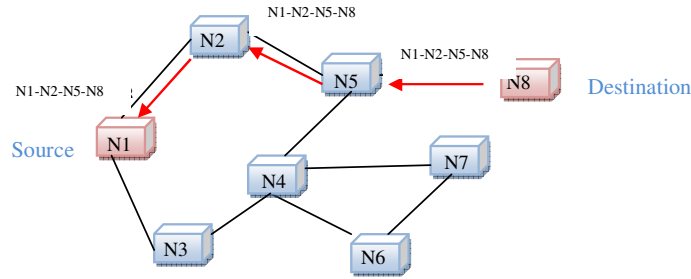

Figure 10. DSR: route reply

Route Discovery with AODV is similar, but also broadcast from the source client to send a Route Request, the difference is, Route Request after one for each hop, this hop of the ID will be recorded in the Route Request a Route Record, the way, When the Route Request reaches destination, they will have all the nodes in the path of information, destination in many elected a Request where the best path, according to Route Record to send a Route Reply back to source, source will be recorded in the route reply stored inside the route record in the routing table, then all should be sent to the destination of the packet will have the route record on the inside, only the source path need to have this information, source and destination paths between other nodes in the packet as long as the view inside the route record and then forward you can go out without having to re-select the path.

## 11.8 Temporally Ordered Routing Algorithm (TORA)

Temporally Ordered Routing Algorithm [40, 41] (TORA). concept of link reversal will be used in MANETS become a kind of decentralized routing algorithms, from the beginning need to route the data sender to the destination number can be used to identify the direction of the path, suitable for highly dynamic environment. TORA is that when there is link the characteristics of the changes, produced by the adjacent node control message is confined to changes occurring at the vicinity of its operation is divided into three steps: Route Creation, Route Maintenance, Route Erasure. In general the process of establishing the route, the network where each node is assigned a "height", and to form a destination as the root of a directed acyclic graph (DAG), the adjacent nodes according to height, all the link will be was designated as upstream or downstream. When network topology changes, DAG was forced to re-establish, when to change the link to the upstream of the upstream node in the direction of height is responsible for the update, until the date back to the source, this update process can source that topology change, while not addressed to destruction of the other route. Topology change from the source to be updated at the beginning to the manner and distance vector routing prone to the same count-to-infinity problems, but this is only a temporary instability.

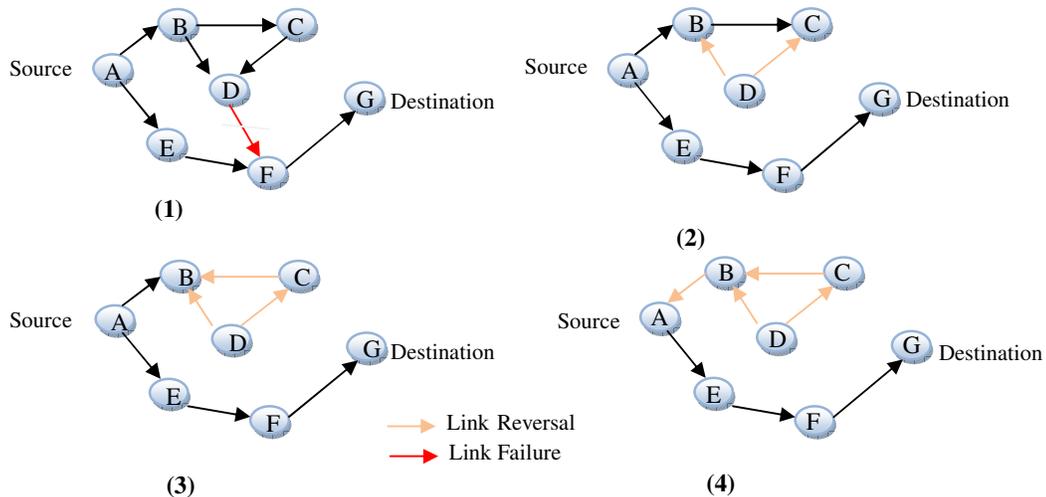

Figure 11. TORA: route maintenance.

### 11.9 Cluster Based Routing protocol (CBRP)

In Cluster Based Routing protocol (CBRP) [60], all the nodes are separated into clusters. In order to arrange the cluster, the following algorithm is used. When a node comes up, it will go into the "undecided" state and broadcasts a Hello message. When a cluster-head gets this hello message it will reacts with a triggered hello message immediately. When the undecided node gets this message it changes its state to "member". If the undecided node times out, then it will turn as a cluster-head. If it has a bi-directional link to some neighbour, otherwise it remains in undecided state and repeats the procedure all over again. Cluster-heads are changed as occasionally as possible.

Each node maintains a neighbour table. For each neighbour, the neighbor table of a node contains the status of the link and the state of the neighbour (clusterhead or member). A cluster head maintains information about the members of its cluster and also maintains a cluster adjacency table that contains information about the neighboring clusters. For each neighbor cluster, the table has entry that contains the gateway through which the cluster can be reached and the information about the cluster head. When a source has to send data to destination, it floods route request packets (but only to the neighboring cluster-heads). While receiving the request, a cluster head will checks whether the destination is in its cluster. If yes, then it sends the request directly to the destination or it will send to all its adjacent cluster-heads. The cluster-heads address is documented in the packet. Then a cluster-head will discards a request packet that has been detected. When the destination receives the request packet, it will reply with the route that had been recorded in the request packet. If the source does not receive a reply within a stipulated time period, it will backs off exponentially before trying to send out another route request.

In CBRP, routing is done by using the source routing. It also uses the route shortening when receiving a source route packet. The node will attempt to find the farthest node in the route and sends the packet to that node and therefore reducing the route. In the process of transferring the packet, if a node detected a broken link, it will send back an error message to the source and then uses a local repair mechanism. In local repair mechanism, if a node detected that the next hop is inaccessible, it will investigate to see if the next hop can be reached through any of its neighbor. If any one of it works, the packet can be sent out over the repaired path.

### 11.10 Associativity Based Routing (ABR)

Associativity Based Routing [42, 43, 44] is primarily designed to focus on the MANET where links between nodes, unstable relations, and therefore uses the concept of associativity stability, is used to indicate a node relative to the adjacent node, link stability. ABR adjacent nodes on a regular basis by forming the beacon to indicate its presence, when a node n received from the adjacent nodes coming beacon, n will be updated on the associativity table, in which each adjacent node in the associativity table where the record is called associativity tick, said that the node relative to the n degree of stability.  The main objective of ABR is to provide the most stable path between nodes, the path establishment process is as follows:  When the node need to go the path of a node, broadcasting a BQ (broadcast query) message, the node will receive the query and its associativity table its own address on its neighboring nodes, together with the associativity ticks along with BQ continues to broadcast out, downstream node d will be its upstream node u where the information recorded in the BQ removed, leaving only, and d the associativity tick, that is, u, d, link stability. BQ way out radio, arrived at destination, it is already recorded on the path from the source to the destination of all associativity ticks, destination based on these information will be added to the total associativity ticks can be obtained respectively, the stability of each path, destination and so choose the most the appropriate path (most stable), and then along the path to the source terminal send REPLY, along the route through the nodes in its routing table information on the establishment of this path.  ABR also designed the path when the link failure, when the reconstruction method. When the source moves, on the re-BQ-REPLY to the above-mentioned steps, if it is a path of nodes within the half-way move, then the mobile node's upstream node u is necessary to undertake Local Query (LQ [H]), is a restricted hop the number of BQ, which is the path to the reconstruction process in order to limit the topology near the approach to change, if one had not received within the time frame REPLY, u should send up Route Notification (RN), requirement of a node sending LQ [H], if not all the way up the success of the reconstruction path (not received from the destination of the REPLY), the midpoint of the path from source to destination is no longer the beginning LQ, but directly tell the source to re-BQ.

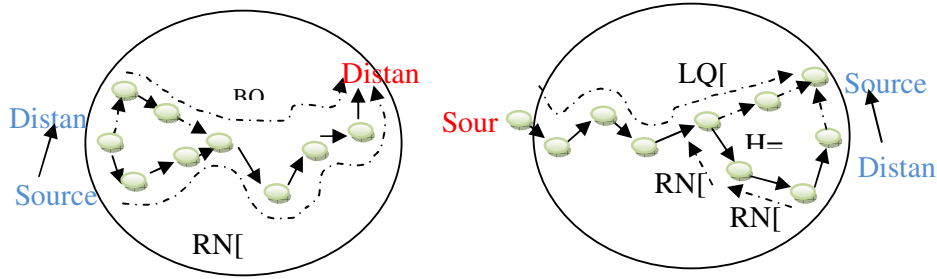

Figure 12. ABR: route maintenance Route maintenance for a source move. (B) Router maintenance for a destination move.

## 11.11 Signal Stability Routing (SSR)

Signal Stability Routing and ABR also added a link to the consideration of stability, is divided into Dynamic Routing Protocol with Static Routing Protocol in two parts. DRP and ABR, as the use of adjacent nodes in each other to define the beacon transmission links are stable, but the DRP only record is strong or weak, that is, a qualitative classification of links, rather than the ABR quantified associatively tick down the value. SRP to use the information obtained by DRP, in the path to the establishment of the process, requires that each downstream node only in the Route request time from the strong link to continue broadcasting route request, select the first to reach the final destination of the request, along the source-side reply to establish a path, so SSR can establish a strong link in the shortest possible, and on the best path.

## 11.12 Core-Extraction Distributed Ad hoc Routing algorithm (CEDAR)

Core-Extraction Distributed Ad hoc Routing algorithm with the above in several different parts of routing protocol is the core use [45]. The original Core is used in the Core Based Tree (CBT), the used to indicate where one or more network nodes are given a special feature may be a relay point for all paths, or the management of certain special features exclusive, In CEDAR, MANET where some of the nodes have been selected for storage of local area link state, and is responsible for calculating the node within the region and select a path.

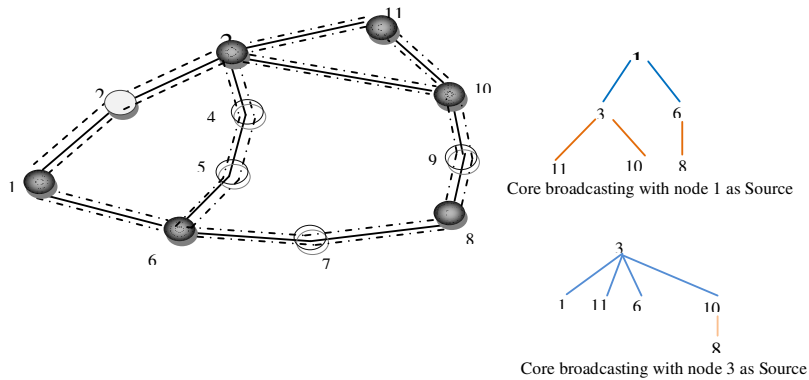

Figure 13. CEDAR: core broadcast

When the source (s) with a destination (d) create a link, first send a request to dom (s), that is, s region of the dominator, dom (s) the destination of the message broadcast out to build a dom (s ) as the starting point of the core path, when the request sent to dom (d) post, dom (d) will conduct reply, at this time dom (s) to dom (d) the path has been established, while dom (d) also, in its local link state to find the best path to d, so the path from s to d, even if the establishment. CEDAR stressing that by storing the relevant Quality of service in the link state information, such as bandwidth, making CEDAR become a Quality of service (QoS) routing; In addition, the definition of a group of network core, simplify the routing process, makes the web of the other nodes do not need to participating in a particular path routing.

### 11.13 Zone Routing Protocol (ZRP)

Zone Routing Protocol combines the path to the establishment of two kinds of reactive and proactive way [46] on the one hand enables the network to keep the record inside a node near the node routing information when a node wants to communicate with neighboring nodes in the path when you can get immediate information, but if you want to, and distant node links, only allow web of a small number of nodes involved in routing.

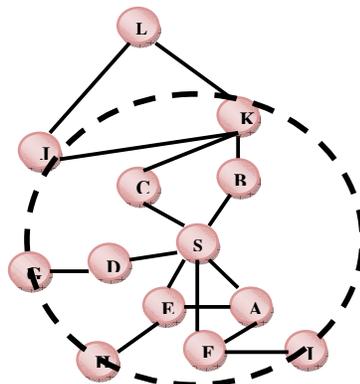

Figure 14. ZRP: a routing zone with radius = 2 (hops)

ZRP defines a node n of neighbors within a hop distance of all the other nodes, as n, routing zone, since each node has its own different adjacent nodes, so ZRP, the entire Internet is in many overlapping routing zone. In the same zone where the routing is called intrazone routing, mainly using the same zone of nodes exchange distance vector, is a form of distance vector routing, but all the distance vector are limited to the size of the zone.

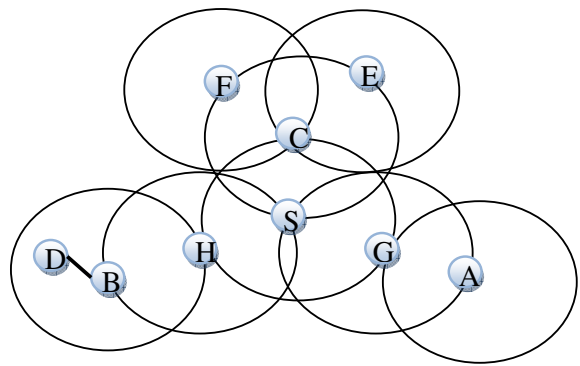

Figure 15. ZRP: interzone operation

When a source is not looking for a destination within its routing zone, when, source use Bordercasting, that is, the route query sent to the source of all the nodes within the routing zone, each node checks the query's destination is within its routing zone if not, continues to make Bordercasting, if there is, you can reply, said that it has found the path can be connected to the destination. Such a process across a number of zone, so called interzone routing.

And now we will shows the comparison between Table Driven, Demand Driven and Hybrid in Table 1, and then we show in table 2 the Table Driven for three kind of protocols such as WRP, CGSR, DSDV and comparison between them, Demand Driven (On-Demand) with six type of protocols such as TORA, DSR, AODV, ABR, CEDAR and SSR and comparison between them shows in table 3.

Table 1. Illustrates Comparison between Table Driven, Demand Driven and Hybrid

|  | Table Driven (Proactive) | Demand Driven (On-Demand) (Reactive) | Hybrid |
|---|---|---|---|
| Routing Protocols | DSDV,CGSR,WRP | AODV,DSR,TORA,ABR,SSR,CEDAR | ZRP |
| Route acquisition delay | Lower | Higher | Lower for Intra-zone; Higher for Inter-zone |
| Control overhead | High | Low | Medium |
| Power requirement | High | Low | Medium |
| Bandwidth requirement | High | Low | Medium |

Table 2 .Shows the Table-Driven for the three kinds of protocols and comparison between them

| Table Driven | CGSR | WRP | DSDV |
|---|---|---|---|
| Routing philosophy | Hierarchical | Flat | Flat |
| Loop-free | Yes | Yes, but not instantaneous | Yes |
| Number of required tables | 2 | 4 | 2 |
| Frequency of update transmissions | Periodically | Periodically and as needed | Periodically and as needed |
| Updates transmitted to | Neighbors and cluster head | Neighbors | Neighbors |
| Utilize hello messages | No | Yes | Yes |
| Critical nodes | Cluster head | No | No |

Table 3 . Shows the Demand Driven (On-Demand) with six types of protocols and comparison between them

| On-Demand | TORA | DSR | AODV | ABR | CEDAR | SSR |
|---|---|---|---|---|---|---|
| Overall complexity | High | Medium | Medium | High | High | High |
| Overhead | Medium | Medium | Low | High | High | High |
| Routing philosophy | Flat | Flat | Flat | Flat | Core-Extracted | Flat |
| Loop Free | Yes | Yes | Yes | Yes | Yes | Yes |
| Multicast capability | No* | No | Yes | No | No* | No |
| Beaconing requirements | No | No | No | Yes | Yes | Yes |
| Multiple route support | Yes | Yes | No | No | No | No |

| Routes maintained in | Route table | Route cache | Route table | Route table | Route table | Route table |
|---|---|---|---|---|---|---|
| Route reconfiguration methodology | Link reversal | Erase route; notify source | Erase route; notify source | Localized broadcast query | Dynamic route re-compute; notify source | Erase route; notify source |
| Routing metric | Shortest path | Shortest path | Freshest & shortest path | Associatively & shortest path & others | Shortest and widest | Associatively & stability |

Table 4. Shows compare the main characteristics of existing multipath routing protocols

| | AODV | DSR | CBRP | DSDV | WRP | GSR | FSR |
|---|---|---|---|---|---|---|---|
| Routing Category | Reactive | Reactive | Reactive | Proactive | Proactive | Proactive | Proactive |
| TTL Limitation | Yas | Yas | Yas | No | No | No | No |
| Flood Control | No | No | No | Yas | Yas | Yas | Yas |
| QoS Support | No | No | No | No | No | No | No |
| Periodic Update | No | No | No | Yas | Yas | Yas | Yas |
| Power Management | No | No | No | No | No | No | No |
| Multicast Support | yas | no | No | No | No | No | No |
| Beaconing | Yas | Yas | Yas | Yas | Yas | Yas | Yas |
| Security Support | No | No | No | No | No | No | No |

## 12.    Quality of Service

With the rapid development of Internet technology, when people for the Best effort service is no longer satisfied, how to get more bandwidth, how to reduce the mistakes, how to reduce the delay phenomenon, making Quality of Service (QoS) related research, including the Integrated Service (RSVP), Differentiated Service, etc., has become an important research topic. In the above-mentioned several agreements, most of them are made in the last two years, only for the basic mode of operation be defined, there is no consideration of QoS, only the ABR (Associativity Based Routing), SSR (Signal Stability Routing) and CEDAR (Core -Extracted Distributed Ad hoc Routing) and so there are three kinds of QoS-related functions.

By ABR, for example, ABR defined by the concept of associativity is that QoS can be used to indicate a link between adjacent nodes stability, while the adjacent node in the exchange of messages, you can also Bandwidth, Delay and other conditions to join, this way then when you select a path, you can have more choices, but also can do according to the different applications of different considerations to select the most appropriate path may be to ensure a minimum bandwidth that can be used, or between two points of a finite delay. However, in the MANET, the network patterns change at any time, each node may change at any time position, that is, each node is the relationship with the adjacent node may change at any time, therefore, means that the need to provide QoS dependent on regular Beaconing, so that each node to master the situation around in order to provide effective QoS information. Beaconing make the overhead on the network increased, when the node mobility to improve even when the general information that may affect the transmission, which will be in the Ad Hoc Network to provide QoS, the biggest problem.

## 13.    Application in Ad hoc Networks

There are a lot of potential applications applied on the Ad hoc networks and to support the Ad hoc Network Model to create a simple Ad Hoc Network, and that application such as the European telecommunications standard institute (ETSI) also the HIPERLAN/2 standard[47,48], IEEE 802.11 wireless LAN standard family [49,50,51,52] and Bluetooth [53] the Ad Hoc Network are very important area in this time and very useful for the military (battlefield) and for the disasters (flood, fire and earthquake and so on),meetings or conventions in which people wish to quickly share information [54].and then use it in the emergency search-and-rescue operations, recovery, home networking ,as we will discuss that in the next table.

Nowadays, Ad Hoc Network became so important in our circle life, because can be applied anywhere where there is little or without communication infrastructure or may be the existing infrastructure is expensive to use. Also the Ad Hoc Networking allows to nodes or devices to keep the connections to the network for as long as it's easy to add and to remove to the end of the network. And there are a lot of varieties of applications for the Mobile Ad hoc Networks, ranging large scale such as Dynamic Network and Mobile and small fixed-constrained energy sources. As well as legacy applications that move from the traditional environment to the Ad Hoc infrastructure environments, a great deal of new services can and will be generated for the new environment, finally as the result the mobile Ad Hoc Network is the important technique for the future and to became for the fourth generation (4G), and the main goals for that to provide propagation the computer environments, that support the users to achieved the tasks to get the information and communicate at anytime, anyplace and from any nodes or devices[55]. And now we will present some of these practical applications as been arranged in table 5, and then we will discuss some of these application.

Table 5. Shows some of application for the Ad Hoc Networks.

| Applications | The Possible Service of Ad Hoc Networks |
|---|---|
| Tactical networks | •Military communication.<br>•Military operations.<br>• in the battlefields. |
| Emergency services | •Search and rescue operations in the desert and in the mountain and so on.<br>• Replacement of fixed infrastructure in case of environmental disasters<br>• Policing<br>•fire fighting<br>• Supporting doctors and nurses in hospitals |
| Coverage extension | • Extending cellular network access<br>• Linking up with the Internet, intranets, and so on. |
| Sensor networks | • Inside the home: smart sensors and actuators embedded in consumer electronics.<br>• Body area networks (BAN)<br>• Data tracking of environmental conditions, animal movements, chemical/biological detection |
| Education | •Universities and campus settings<br>• Classrooms<br>• Ad hoc Network when they make a meetings or lectures |
| Education | •Multi-user games<br>• Wireless P2P networking<br>• Outdoor Internet access<br>• Robotic pets<br>• Theme parks |
| Home and enterprise | •Using the wireless networking in Home or office.<br>• Conferences, meeting rooms |

| networks | • Personal area networks<br>• Personal networks. |
|---|---|
| Context aware services | • Follow-on services: call-forwarding, mobile workspace<br>• Information services: location specific services, time dependent services<br>• Infotainment: touristic information |
| Commercial and civilian environments | • E-commerce: electronic payments anytime and anywhere<br>• Business: dynamic database access, mobile offices<br>• Vehicular services: road or accident guidance, transmission of road and weather conditions, taxi cab network, inter-vehicle networks<br>• Sports stadiums, trade fairs, shopping malls and so on.<br>• Networks of visitors inside the airports. |

These are a lot of applications on Ad Hoc Networks as we saw in the last part and also in the table 5 provide an overview of present and future Mobile Ad Hoc Networks Applications. However, now we will discuss more about some of this Applications such as Tactical Networks (military battlefield), Home and enterprise Network (personal area network (PAN)) and

**13.1  Military battlefield**. Military equipment currently is equipped with the state of the art computer equipment. Ad hoc networking help the military with the commonplace network technology to maintain information network between military personnel's, vehicles, and military information head quarters. The basic techniques of ad hoc network originated from this field.

**13.2  Commercial sector.** Ad hoc network can be applied in emergency or rescue operations for disaster relief efforts for example in fire, flood, or earthquake and so on. Emergency rescue operations will go to places where communications are impermissible. Therefore proper infrastructure and rapid deployment of a communication network is badly needed. Information is relayed from one rescue team member to another over a small handheld device. Other commercial application includes for instance ship to ship Ad Hoc Mobile communication and so on.

**13.3   Local level**. Ad hoc networks can autonomously link immediate and temporary multimedia network by using notebook or palmtop computers to distribute and allocate information among conference or classroom participants. Besides, it can also be applied for home networks where devices can be link. Another example includes taxicab, sports stadium, boat and small aircraft.

**13.4  Personal Area Network (PAN).** Short-range MANET can simplify the intercommunication between a lot of mobile devices such as a PDA, a laptop, and a cellular phone and there are a lot of new devices in this for MANETs. Wired cables can easily be replaced with wireless connections. Ad hoc network enhances the access to the Internet or other networks by means of Wireless LAN (WLAN), GPRS, and UMTS. The PAN is an upcoming application field of MANET for the future computing technology.

## 14.  CONCLUSIONS

In this paper we presented an exhaustive survey about the Mobile Ad Hoc Network (MANET) we distinct the characteristics of traditional wired networks, wireless ad hoc networks, wireless mobile approaches and types of ad hoc network as well as all the existing ad hoc protocols, and we comparison between the different papers, most of its conclusions pointed to a phenomenon, not a routing protocol can adapt to all environments, whether it is Table-Driven, On-Demand or a mixture of two kinds, are limited by the network characteristics; even though the same part of the Agreement On-Demand also due to the differences in the mode of operation applicable to different types of network. Also we discussed in this paper the relevant Ad Hoc Network on a multicast (Multicasting), Applications on Ad Hoc Networks, QoS and other topics will be able to see the latest research results, can be expected is that the Ad Hoc Network needs and applications will start to appear in recent years, Ad Hoc Network-related research have become the current Internet trends One of the most anticipated technology.

## ACKNOWLEDGMENT

Special thank and recognition go to my advisor, Associate Professor. Dr. Putra Sumari, who guided me through this reserech, inspired and motivated me. Last but not least, the researchers would like to thank the University Sains Malaysia USM for supporting this research.

**Authors**

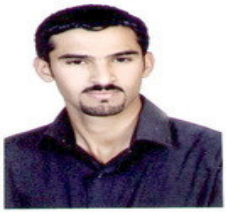
**Saleh Ali K. Al-Omari** Obtained his Bachelor degree in Computer Science from Jerash University, Jordan in 2004-2005 and Master degree in Computer Science from University Sains Malaysia, Penang ,Malaysia in 2007. Currently, He is a PhD candidate at the School of Computer Science, University Sains Malaysia, Penang, His main research interest now on Video on Demand (VoD) over Heterogeneous Mobile Ad Hoc networks (MANETs).

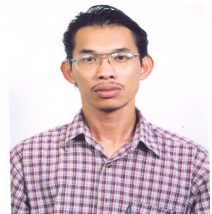
**Putra Sumari** is currently working as Assistant Professor of School of Computer Science, Universiti Sains Malaysia, Penang in 2009. Assoc.Prof Dr.Putra received his MSc and PhD in 1997 and 2000 from Liverpool University, England. Currently, he is a lecturer at the School of Computer Science, Universiti Sains Malaysia, Penang. He is the head of the Multimedia Computing Research Group, School of Computer Science, USM. He is Member of ACM and IEEE, Program Committee and reviewer of several International Conference on Information and Communication Technology (ICT), Committee of Malaysian ISO Standard Working Group on Software Engineering Practice, Chairman of Industrial Training Program School of Computer Science USM, Advisor of Master in Multimedia Education Program, UPSI, and Perak.